# Magic Number Theory of Superconducting Proximity Effects and Wigner Delay Times in Graphene-Like Molecules

P. Rakyta[#], A. Alanazy[*], A. Kormányos[#], Z. Tajkov[+], G. Kukucska[+], J. Koltai[+], S. Sangtarash[*], H. Sadeghi[*], J. Cserti[#] and C.J. Lambert[*]

[#]Dept. of Physics of Complex Systems, Eötvös Loránd University, Budapest, Pázmány P. s. 1/A, Hungary

[*]Dept. of Physics, Lancaster University, Lancaster, LA1 4YB, United Kingdom.

[+]Dept. of Biological Physics, Eötvös Loránd University, Budapest, Pázmány P. s. 1/A, Hungary

**ABSTRACT**

When a single molecule is connected to external electrodes by linker groups, the connectivity of the linkers to the molecular core can be controlled to atomic precision by appropriate chemical synthesis. Recently, the connectivity dependence of the electrical conductance and Seebeck coefficient of single molecules has been investigated both theoretically and experimentally. Here we study the connectivity dependence of the Wigner delay time of single-molecule junctions and the connectivity dependence of su-perconducting proximity effects, which occur when the external electrodes are replaced by superconductors. Although absolute values of transport properties depend on complex and often uncontrolled details of the coupling between the molecule and electrodes, we demonstrate that ratios of transport properties can be predicted using tables of 'magic numbers,' which capture the connectivity dependence of superconducting proximity effects and Wigner delay times within molecules. These numbers are calculated easily, without the need for large-scale computations. For normal-molecule-superconducting junctions, we find that the electrical conductance is proportional to the fourth power of their magic numbers, whereas for superconducting-molecule-superconducting junctions, the critical current is proportional to the square of their magic numbers. For more conventional normal-molecule-normal junctions, we demonstrate that delay time ratios can be obtained from products of magic number tables.

Corresponding authors: Colin Lambert: Email c.lambert@lancaster.ac.uk; Jozsef Cserti: Email cserti@elte.hu

1. **INTRODUCTION**

During the past decade, experimental and theoretical studies of single molecules attached to metallic electrodes have demonstrated that room-temperature electron transport is controlled by quantum interference (QI) within the core of the molecule[1-20]. These studies provide tremendous insight into the mechanisms leading to efficient charge transport, but they ignore key aspects of quantum mechanical phase. For example, such junctions are often described using the Landauer formula $G = G_0 T(E_F)$, where $G_0 = \frac{2e^2}{h}$ is the quantum of conductance and $E_F$ is the Fermi energy of the electrodes. In this expression $T(E)$ is the transmission coefficient describing the probability that an electron of energy $E$ can pass through the junction from one electrode to the other and for single-channel leads, is related to the transmission amplitude $t(E)$ by $T(E) = |t(E)|^2$, where $t(E)$ is a complex number of the form $t(E) = |t(E)|e^{i\theta(E)}$.

Clearly the phase $\theta(E)$ of the transmission amplitude plays no role when computing $T(E)$, even though $T(E)$ is a result of interference from different transport channels within a molecular junction.

The aim of the present paper is to examine examples of molecular-scale transport in which phase plays a crucial role and to discuss aspects of molecular-scale electron transport when one or more electrodes are superconducting. An example of such a structure is a normal-electrode/molecule/superconducting-electrode junction (denoted N-M-S), where the electrical conductance is proportional to the Andreev reflection coefficient of the junction. A second example is a molecular-scale Josephson junctions (denoted S-M-S') formed by placing a single molecule between two superconducting electrodes S and S'. In this case the dc electrical current is driven by the phase difference between the order parameters of the two superconductors. A third example occurs in N-M-SS' junctions, whose Andreev refection coefficient is an oscillatory function of the phase difference between the two superconducting contacts S, S'. These considerations are motivated by the recent interest in superconducting properties of molecular scale junctions[21,22,23,24]. Secondly, we study examples where the phase θ(E) of the transmission amplitude plays a crucial role in normal N-M-N junctions. In this case, the phase $\theta(E)$ is related to the Wigner delay time, which characterises the time taken for an electron to pass through a single-molecule junction formed from normal electrodes.

2. **METHODS**

To illustrate how these phase-dependent phenomena can be predicted using magic number theory, Figure 1 shows two examples of molecules with a graphene-like anthanthrene core, connected via triple bonds and pyridyl anchor groups to gold electrodes. The anthanthrene core (represented by a lattice of 6 hexagons) of molecule **1** and the anthanthrene core of molecule **2** are connected differently to the triple bonds.

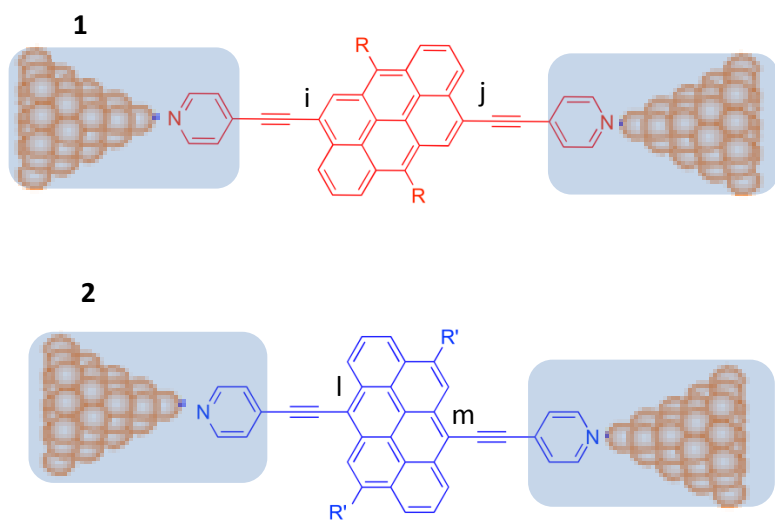

**Figure 1**: Examples of molecules with anthanthrene cores, connected via triple bonds and pyridyl anchor groups to the tips of gold electrodes, which in turn connect to crystalline gold leads (not shown). Molecule **1** has a connectivity *i-j* and electrical conductance $\sigma_{ij}$, while molecule **2** has a connectivity *l-m* and electrical conductance $\sigma_{lm}$.

In a typical experiment using mechanically controlled break junctions or STM break junctions [13-18,21], fluctuations and uncertainties in the coupling to normal-metallic electrodes are dealt with by measuring the conductance of such molecules many thousands of times and reporting the statistically-most-probable electrical conductance. If $\sigma_{ij}$ is the statistically-most-probable conductance of a molecule such as **1** (see Figure 1), with connectivity *i-j* and $\sigma_{lm}$ is the corresponding conductance of a molecule such as **2** (see Figure 1), with connectivity *l-m*, then it was recently predicted theoretically and demonstrated experimentally[2,25,26] that for polyaromatic hydrocarbons such as anthanthrene, the statistically-most-probable conductance ratio $\sigma_{ij}/\sigma_{lm}$ is independent of the coupling to the electrodes and could be obtained from tables of "magic numbers." If $M_{ij}$ ($M_{lm}$) is the magic number corresponding to connectivity *i-j* (*l-m*), then this "magic ratio theory" predicts

$$\frac{\sigma_{ij}}{\sigma_{lm}} = \left(\frac{M_{ij}}{M_{lm}}\right)^2 \qquad (1)$$

From a conceptual viewpoint, magic ratio theory views the shaded regions in Fig. 1 as "compound electrodes", comprising both the anchor groups and gold electrodes, and focuses attention on the contribution from the core alone. As discussed in[28], the validity of Eq. (1) rests on the key foundational concepts of weak coupling, locality, connectivity, mid-gap transport, phase coherence and connectivity-independent statistics. When these conditions apply, the complex and often uncontrolled contributions from electrodes and electrode-molecule coupling cancel in conductance ratios and therefore a theory of conductance ratios can be developed by focussing on the contribution from molecular cores alone.

The term "weak coupling" means that the central aromatic subunit such as anthanthrene should be weakly coupled to the anchor groups via spacers such as acetylene, as shown in Fig. 1. Weak coupling means that the level broadening $\Gamma$ and the self energy $\Sigma$ of the HOMO and LUMO should be small compared with the HOMO-LUMO gap $E_{HL}$. Any corrections will then be of order $\Gamma/E_{HL}$ or $\Sigma/E_{HL}$, which means that such terms can be ignored, provided the Fermi energy lies within the gap. Clearly a central condition for the applicability of the Landauer formula and therefore magic-number theory is that the molecular junction is described by a time independent mean-field Hamiltonian. Coulomb interactions can be included in such a Hamiltonian, at the level of a self-consistent mean field description such as Hartree, Hartree-Fock or DFT. The concept of 'mid-gap transport' is recognition of the fact that unless a molecular junction is externally gated by an electrochemical environment or an electrostatic gate, charge transfer between the electrodes and molecule ensures that the energy levels adjust such that the Fermi energy $E_F$ of the electrodes is located in the vicinity of the centre of the HOMO-LUMO gap and therefore transport takes place in the co-tunnelling regime. In other words, transport is usually 'off-resonance' and the energy of electrons passing through the core does not coincide with an energy level of the molecule. Taken together, these conditions ensure that when computing the Green's function of the core, the contribution of the electrodes can be ignored. The concept of 'phase coherence' recognises that in this co-tunnelling regime, the phase of electrons is usually preserved as they pass through a molecule and therefore transport is controlled by QI. 'Locality' means that when a current flows through an aromatic subunit, the points of entry and exit are localised in space. For example, in molecule **1** (see Figure 1), the current enters at a particular atom *i* and exits at a particular atom *j*. The concept of 'connectivity' recognises that through chemical design and synthesis, spacers can be attached to different parts of a central subunit with atomic accuracy and therefore it is of interest to examine how the flow of electricity depends on the choice of connectivity to the central subunit. The condition of "connectivity-independent statistics" means that the statistics of the coupling between the anchor groups and electrodes should be independent of the coupling to the aromatic core. To be more precise, we note that in an experimental measurement of single-molecule conductance using for example a mechanically-controlled break junction, many thousands of measurements are made and a histogram of logarithmic conductances is constructed. This statistical variation arises from variability in the electrode geometry and in the binding conformaton to the electrodes of terminal atoms such as the nitrogens in figure 1. The assumption of "connectivity-independent statistics" means that this variability is the same for the two different connectivities of figure 1. When each of these conditions applies, it can be shown[2,25,26] that in the presence of normal-metallic electrodes, the most probable electrical conductance corresponding to connectivity *i-j* is proportional to $|g_{ij}(E_\mathrm{F})|^2$ where $g_{ij}(E_\mathrm{F})$ is the Green's function of the isolated core alone, evaluated at the Fermi energy of the electrodes. In the absence of time-reversal symmetry breaking, $g_{ij}(E_\mathrm{F})$ is a real number. Since only conductance ratios are of interest, we define magic numbers by

$$M_{ij} = A g_{ij}(E_\text{F}) \qquad (2),$$

where *A* is an arbitrary constant of proportionality, chosen to simplify magic number tables and which cancels in Eq. (1). Magic ratio theory represents an important step forward, because apart from the Fermi energy $E_\text{F}$, no information about the electrodes is required. The question we address below is how is the theory modified in the presence of superconducting electrodes and how can the theory be extended to describe Wigner delay times?

In the presence of normal-metallic electrodes, many papers discuss the conditions for *destructive* quantum interference (DQI), for which $M_{ij} \approx 0$ [9,18,27,29-33]. On the other hand, magic ratio theory aims to describe *constructive* quantum interference (CQI), for which $M_{ij}$ may take a variety of non-zero values. If $H$ is the non-interacting Hamiltonian of the core, then since $g(E_\text{F}) = (E_\text{F} - H)^{-1}$, the magic number table is obtained from a matrix inversion, whose size and complexity reflects the level of detail contained in $H$. The quantities $M_{ij}$ were termed "magic" [2,25,26], because even a simple theory based on connectivity alone yields values, which are in remarkable agreement with experiment [25]. For example, for molecule **1** (see Figure 1), the prediction was $M_{ij} = -1$, whereas for molecule **2**, $M_{lm} = -9$ and therefore the electrical conductance of molecule **2** was predicted to be 81 times higher than that of **1**, which is close to the measured value of 79. This large ratio is a clear manifestation of quantum interference (QI), since such a change in connectivity to a classical resistive network would yield only a small change in conductance. To obtain the above values for $M_{ij}$ and $M_{lm}$, the Hamiltonian $H$ was chosen to be

$$H = \begin{pmatrix} 0 & C \\ C^t & 0 \end{pmatrix} \qquad (3),$$

where the connectivity matrix $C$ of anthanthrene is shown in Fig. 2. In other words, each element $H_{ij}$ was chosen to be -1 if $i, j$ are nearest neighbours or zero otherwise and since anthanthrene is represented by the bipartite lattice in which odd numbered sites are connected to even numbered sites only, $H$ is block off-diagonal. The corresponding core Green's function evaluated at the gap centre $E_F = 0$ is therefore obtained from a simple matrix inversion $g(0) = -H^{-1}$. Since $H$ and therefore $-H^{-1}$ are block off-diagonal, this yields $M = \begin{pmatrix} 0 & \bar{M}^t \\ \bar{M} & 0 \end{pmatrix} \propto g(0)$, where $M$ is the magic number table of the polycyclic aromatic hydrocarbons (PAHs) core. The connectivity matrix $C$ and off-diagonal block of the magic number table $M$ for anthanthrene are shown in Figure 2b and c respectively. As noted above, for molecule **1** (see Figure 1), with connectivity 22-9, $M_{22,9} = -1$, whereas for molecule **2**, with connectivity 12-3, $M_{12,3} = -9$.

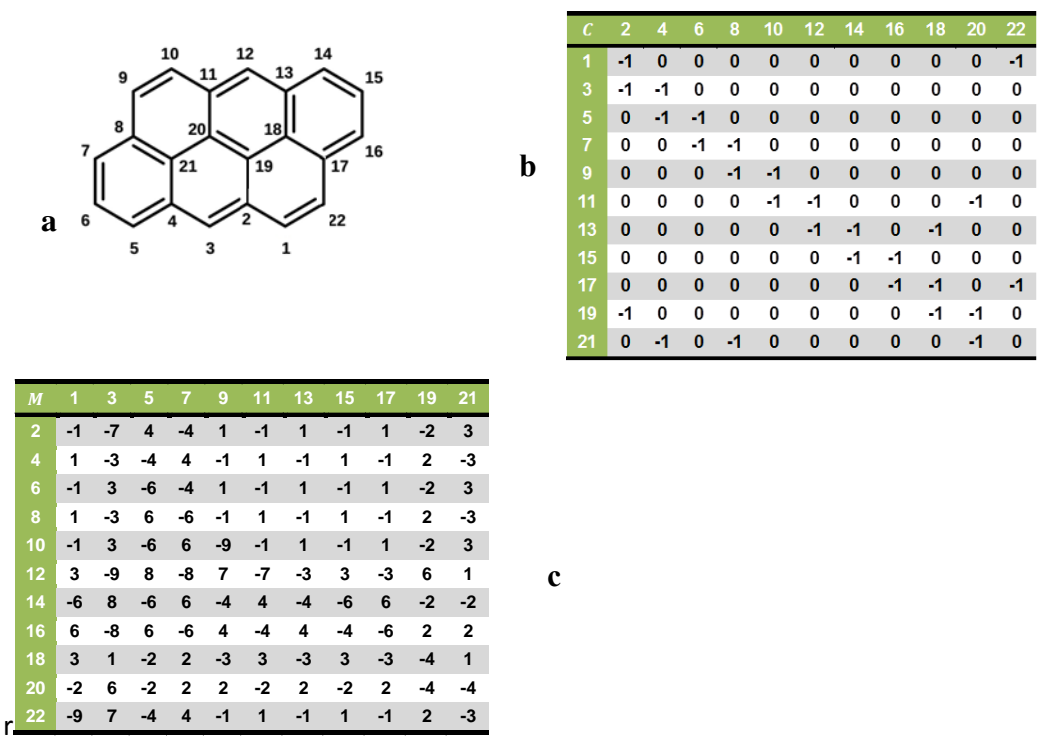

**Figure 2:** (a) The anthanthrene cores numbering system. (b) The connectivity table $C$. (c) The off-diagonal block of the magic number table corresponding to the anthanthrene lattice. In this example, equation (2) takes the form $M_{ij} = 10 g_{ij}(0)$.

Magic number tables such as Figure *2*c are extremely useful, since they facilitate the identification of molecules with desirable conductances for future synthesis. Conceptually, tables obtained from Hamiltonians such as (3) are also of interest, since they capture the contribution from intra-core connectivity alone (via the matrix $C$, comprising -1's or zeros), while avoiding the complexities of chemistry. The fact that magic number theory predicts experimentally-measured conductance ratios for a range of molecules[25] demonstrates that at least for PAHs measured to date, magic number theory is valid. In what follows, the studied molecules are chosen, because their behaviour in the presence of normal-metal (ie gold) contacts they exhibit a sizeable connectivity dependence in their electrical conductances, as demonstrated experimentally and theoretically in ref [25]. The tight binding parameters describing the molecular core are chosen following the philosophy in refs[2,25], where the aim is to highlight the role of connectivity in determining the transport properties of these molecular cores. For this reason, the hopping integrals $\gamma$ are set to unity and the site energies $\epsilon_0$ are set to zero. In other words, the unit of energy is the hopping integral and the site energy is the energy origin. This means that the Hamiltonian is simply a connectivity matrix and therefore all predicted effects are a result of connectivity alone. Remarkably, as demonstrated in [2,25], this approach yields the experimentally-measured conductance ratios of a range of PAHs.

In the wide band limit, the only other parameter is the coupling between the terminal sites and the electrodes. Our aim is to compute ratios of transport properties corresponding to different connectivities to the electrodes. As shown in refs [2,25], transport ratios do not depend on these couplings, provided they are sufficiently weak. Comparison with experiment in these refs shows that this weak-coupling criterion is satisfied by acetylene linker groups connecting the aromatic core to pyridyl anchor groups, which in turn bind to electrodes, as shown in Fig. 1.

## 3. ANALYTIC RESULTS

For convenience, we first state our main analytic results. Details of the derivations and discussion of the results will be given in later sections and in the Supplementary Information. The first outcome of the study is that for normal-molecule-superconducting (N-M-S) junctions Eq. (1) is replaced by

$$\frac{\sigma_{ij}}{\sigma_{lm}} = \left(\frac{M_{ij}}{M_{lm}}\right)^4, \tag{4}$$

where $i$ and $l$ label the atoms in contact with the normal electrode, while $j$ and $m$ label atoms in contact with the superconducting electrode. Equation (4) shows that ratios of electrical conductances are determined by the fourth power of magic numbers. The fourth power in the formula can be explained via the mechanism of the Andreev reflection (for further details see Sec. 4a). In the case of Josephson junctions formed from superconducting-normal-superconducting (S-M-S) structures, the ratio of their critical currents $I_c^{(ij)}$ and $I_c^{(lm)}$ corresponding to different connectivities is given by

$$\frac{I_c^{(ij)}}{I_c^{(lm)}} = \left(\frac{M_{ij}}{M_{lm}}\right)^2 \tag{5}$$

where $i$ and $l$ label the atoms in contact with one superconducting electrode, while $j$ and $m$ label atoms in contact with the other superconducting electrode. For molecular-scale Andreev interferometers, where a molecule is attached two superconducting contacts and one normal contact (N-M-SS'), the conductance through the normal contact is given by the formula

$$\frac{\sigma_{l,mp}(\varphi_R - \varphi_L)}{\sigma_{l,mp}(0)} = \frac{M_{lm}^4 + M_{lp}^4 + 2M_{lm}^2 M_{lp}^2 \cos(\varphi_R - \varphi_L)}{\left(M_{lm}^2 + M_{lp}^2\right)^2}, \tag{6}$$

where $\varphi_R - \varphi_L$ is the superconducting phase difference between the right and left superconducting electrodes. In equation (6), $l$ labels the atom in contact with the normal electrode, while $p$ ($m$) label the atom in contact with the superconducting electrode S.

According to equation (6) the current through the normal lead is sensitive to the phase difference $\varphi_R - \varphi_L$. As we shall see in Sec. 4c, this phenomenon can be understood as an interference effect between two transport paths. Finally, for N-M-N junctions (see section 5), the ratio of Wigner delay times corresponding to connectivities $i,j$ and $l,m$ is

$$\frac{\tau_{ij}}{\tau_{lm}} = \frac{\tau_{ii} + \tau_{jj}}{\tau_{ll} + \tau_{mm}} \qquad (7),$$

where

$$\tau_{ii} = (M^2)_{ii} \qquad (8)$$

### 4. RESULTS FOR MOLECULAR JUNCTIONS WITH ONE OR MORE SUPERCONDUCTING ELECTRODES

In the presence of superconductivity, electron transport is controlled by both normal electron scattering and Andreev scattering. When the normal region between electrodes is a diffusive metal, one traditionally treats transport using quasi-classical equations[34]. Such an approach is not appropriate for phase-coherent transport through molecular cores, where the arrangement of atoms within the central region is deterministic. Instead, one should use a scattering approach which preserves such atomic-scale details[35,36] by solving the Bogoliubov-de Gennes equation for the Green's function of the junction[37]. In what follows, we consider proximity effects in normal-molecule-superconducting (N-M-S) junctions containing a single superconducting electrode, in normal-molecule-double superconducting (N-M-SS') junctions containing two superconducting electrodes denoted by S and S' and in superconducting-molecule-superconducting (S-M-S') Josephson junctions containing two superconducting electrodes. In junctions containing two superconducting electrodes, time reversal symmetry can be broken by a finite superconducting phase difference $\varphi_R - \varphi_L$ between the two superconductors S and S'. Thus, for S-M-S' junctions, imposition of a superconducting phase difference $\varphi_R - \varphi_L$ would generate a Josephson current even in the absence of an applied voltage[38]. Moreover, in (N-M-SS') junctions the finite phase difference induces an oscillation in the electrical conductance as a periodic function of $\varphi_R - \varphi_L$ [39] We note that the phase difference $\varphi_R - \varphi_L$ can be controlled experimentally[40,41], allowing the measurement of the current phase relation in S-M-S' junctions or the measurement of the phase dependence of the electrical conductance in N-M-SS' junctions. In what follows, all three types of junctions will be considered.

**4a. N-M-S junctions**

In N-M-S junctions during a single scattering process an incoming electron and an Andreev reflected hole passes through the molecule. Thus, the transmission amplitude, in contrast to the normal transport, is proportional to the square of the magic numbers. For instance, if the normal and superconducting electrodes are connected to sites $l$ and $m$, the transmission amplitude would be proportional to $t_{lm} \sim -M_{lm}{}^2 e^{-i\varphi}$ (see the Supporting Information for further details). The second power of $M_{lm}$ arises from the fact that during Andreev reflection, an electron traverses the molecule in one direction and then a hole traverses the molecule in the other. The minus sign originates from the Green function elements related to the propagation of the reflected holes, and in addition the Andreev reflection contributes to the total transmission amplitude with a phase factor $e^{-i\varphi}$, where $\varphi$ is the phase of the superconducting order parameter. However, the conductance is not sensitive to the specific value of this phase factor. Calculating the conductance using the Landauer formula, one can indeed arrives at equation (4). To demonstrate the validity of equation (4) we consider the normal-molecule-superconducting junctions with a pyrene core, shown in Figure 3.

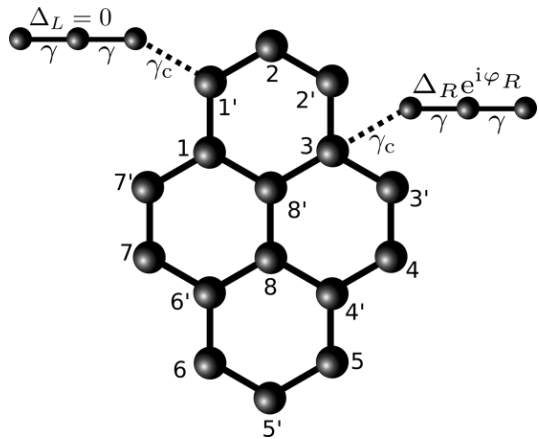

**Figure 3**: Normal-pyrene-superconducting junction with normal electrode coupled to site 1' and with superconducting contact at site 3. γ is the hopping amplitude between atoms, Δ is the pairing potential in the superconducting contact and $γ_c$ is the coupling between the molecule and the superconducting contacts.

In our numerical calculation we used the tight binding model to describe the non-interacting Hamiltonian of the molecule with hopping amplitude γ. Following the reasoning of reference[42], the conductance can be calculated using the normal and Andreev reflection amplitudes by the formula:

$$\sigma_{ij} = \frac{2e^2}{h}\left(N - R_{0,ij} + R_{a,ij}\right) = \frac{4e^2}{h} R_{a,ij}, \tag{9}$$

where $N$ is the number of propagating electron-like channels in the normal lead, $R_{0,ij}$ is the normal and $R_{a,ij}$ is the Andreev reflection coefficient associated with a normal lead connected to site $i$, and with a superconducting lead connected to site $j$ of the molecule. The second equality is valid for low biases $eV \ll |\Delta|$, where Δ is the superconducting pair potential. These reflection coefficients can be calculated via the Green's function theory of reference[43].

Using the numbering convention of the sites given in Figure 3, one can easily show that the magic numbers are non-zero only between non-primed i and primed j' sites. The magic numbers between sites i and j' are shown in Table 1. We assumed that the coupling $γ_c$ between the molecule and the superconducting contacts is weak. Then it is reasonable to take the pairing potential Δ to be non-zero only in the contact.

|   | 1' | 2' | 3' | 4' | 5' | 6' | 7' | 8' |
|---|---|---|---|---|---|---|---|---|
| 1 | 3 | -3 | 1 | -1 | 1 | -1 | 1 | 2 |
| 2 | 3 | 3 | -1 | 1 | -1 | 1 | -1 | -2 |
| 3 | -3 | 3 | 1 | -1 | 1 | -1 | 1 | 2 |
| 4 | 3 | -3 | 5 | 1 | -1 | 1 | -1 | -2 |
| 5 | -3 | 3 | -3 | 3 | 3 | -3 | 3 | 0 |
| 6 | 3 | -3 | 3 | -3 | 3 | 3 | -3 | 0 |
| 7 | -3 | 3 | -1 | 1 | -1 | 1 | 5 | -2 |
| 8 | 0 | 0 | -2 | 2 | -2 | 2 | -2 | 2 |

**Table 1**: The magic numbers for pyrene. Use: See Figure 3 for the meaning of indexes i and j'. Magic numbers connecting two sites both labeled by a prime (or both labeled without a prime) are zero.

We calculated numerically the electrical conductances between different pairs of sites $i$ and $i'$. The superconducting pairing potential in our calculations was $|\Delta| = 1$ meV, the hopping amplitude was γ = 2.4 eV and coupling amplitude was $γ_c = 0.45$ γ. The onsite potential was zero at all sites. Our numerical results are summarized in Table *2*. As one can see, our theoretical prediction in Eq. (4) is confirmed; namely the ratio of the calculated conductance between different sites $i$ and $i'$ agrees very well by the fourth power of magic numbers given in Table *1*.

|   | 1'    | 2'    | 3'     | 4'    | 5'    | 6'    | 7'     | 8'    |
|---|-------|-------|--------|-------|-------|-------|--------|-------|
| 1 | 80.98 | 80.98 | 1.00   | 1.00  | 1.00  | 1.00  | 1.00   | 16.00 |
| 2 | 80.98 | 80.98 | 1.00   | 1.00  | 1.00  | 1.00  | 1.00   | 16.00 |
| 3 | 80.98 | 80.98 | 1.00   | 1.00  | 1.00  | 1.00  | 1.00   | 16.00 |
| 4 | 80.98 | 80.98 | 623.99 | 1.00  | 1.00  | 1.00  | 1.00   | 16.00 |
| 5 | 80.98 | 80.98 | 80.98  | 80.98 | 80.98 | 80.98 | 80.98  | 0.00  |
| 6 | 80.98 | 80.98 | 80.98  | 80.98 | 80.98 | 80.98 | 80.98  | 0.00  |
| 7 | 80.98 | 80.98 | 1.00   | 1.00  | 1.00  | 1.00  | 623.99 | 16.00 |
| 8 | 0.00  | 0.00  | 16.00  | 16.00 | 16.00 | 16.00 | 16.00  | 16.00 |

**Table 2:** The calculated conductance (in units of $\sigma_0 = 1.0375 \times 10^{-5}$ $e^2/h$) through a single pyrene molecule between contacting sites i and j' calculated for $\gamma_c = 0.45\ \gamma$. Conductances connecting two sites both labelled by a prime (or both labelled without a prime) are zero.

Note that the values of individual conductances have no significance in table 2, since our only aim is to calculate conductance ratios. The conductances are determined by the arbitrary couplings to the electrodes, which do not affect conductance ratios. Furthermore conductance ratios are independent of the parameters used to define the electrodes. The above value of Δ was chosen, because superconducting gap of common superconductors such as niobium, tantalum and mercury is on the scale of a meV. The value of γ_c was chosen to ensure that the level broadening due to the contacts is small compared to the HOMO-LUMO gap, which is in the experimentally-relevant regime. Just as conductance ratios are independent of the parameters used to define the normal lead, they are also independent of the parameters used to define the superconducting electrode, including the size of the superconducting energy gap.

Magic number theory is a weak coupling theory and eventually, as the coupling to the electrodes increases, there will be a difference between magic number theory and a full tight binding calculation. To illustrate this point and at the same time to show that the deviation is small, we performed the same calculation for $\gamma_c = 0.85\ \gamma$, which is larger than in the previous calculation. The results are given in Table 3. Note that the ratios of the calculated conductance deviate from the fourth power law of magic numbers given in equation (4). However, for $\gamma_c < 0.85\ \gamma$ the deviation from the weak coupling limit remains of order 10 %.

|   | 1'    | 2'    | 3'     | 4'    | 5'    | 6'    | 7'     | 8'    |
|---|-------|-------|--------|-------|-------|-------|--------|-------|
| 1 | 78.35 | 78.35 | 1.00   | 1.00  | 1.00  | 1.00  | 1.00   | 15.90 |
| 2 | 78.35 | 78.35 | 1.00   | 1.00  | 1.00  | 1.00  | 1.00   | 15.90 |
| 3 | 78.35 | 78.35 | 1.00   | 1.00  | 1.00  | 1.00  | 1.00   | 15.90 |
| 4 | 78.35 | 78.35 | 488.57 | 1.00  | 1.00  | 1.00  | 1.00   | 15.90 |
| 5 | 78.35 | 78.35 | 78.35  | 78.35 | 78.35 | 78.35 | 78.35  | 0.00  |
| 6 | 78.35 | 78.35 | 78.35  | 78.35 | 78.35 | 78.35 | 78.35  | 0.00  |
| 7 | 78.35 | 78.35 | 1.00   | 1.00  | 1.00  | 1.00  | 488.57 | 15.90 |
| 8 | 0.00  | 0.00  | 15.90  | 15.90 | 15.90 | 15.90 | 15.90  | 15.90 |

**Table 3**: The calculated conductance (in units of $\sigma_0 = 0.0017\ e^2/h$) through a single pyrene molecule between contacting sites i and j' calculated for $\gamma_c = 0.85\ \gamma$.

### 4b. S-M-S' junctions

In mesoscopic superconductivity it is well known that in phase biased Josephson junctions, at low temperatures, under rather general conditions, the critical current $I_c$ is inversely proportional to the normal state resistance of the junction[44], i.e., $I_c \sim \frac{1}{R_N} \sim \sigma$. Therefore one may expect that in S-M-S' junctions the critical current $I_c^{ij}$ flowing between the superconducting electrodes connected to sites $i$ and $j$ of the molecule is proportional to the transmission amplitude between these points leading to $I_c^{ij} \sim M_{ij}^2$. Equation (5) is a direct consequence of this relation. (A more rigorous derivation is given in the Supporting information.) We now demonstrate the validity of equation (5) by considering the superconducting-normal-superconducting structure with a pyrene core shown in Figure 4. The pyrene molecule was described in the same way as in the case of Andreev reflection. We assumed that the pairing potential Δ is finite only in the superconducting contacts indicated in the Figure *4*.

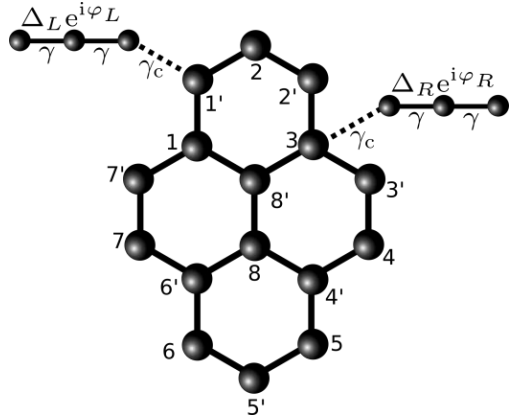

**Figure 4**: Superconducting-normal-superconducting structure with pyrene molecule coupled to two superconducting contacts at site 1' and 3. In the figure γ is the hopping amplitude between atoms, Δ is the pairing potential in the superconducting contact and $\gamma_c$ is the coupling amplitude between the molecule and the superconducting contacts.

Using our numerical method presented in Ref. [45] we calculated the critical current between the different pairs of sites i and j'. We used a superconducting pairing potential |Δ| = 1 meV, a hopping amplitude γ = 2.4 eV and coupling amplitude $\gamma_c$ = 0.45 γ. The onsite potential was zero at all sites. The calculated critical current are summarized in Table *4*. Clearly our theoretical prediction in equation (5) is confirmed, namely the ratio of the calculated critical currents between different sites i and j' agree very well by the squares of the magic numbers given in Table *1*.

|   | 1' | 2' | 3' | 4' | 5' | 6' | 7' | 8' |
|---|---|---|---|---|---|---|---|---|
| 1 | 8.9216 | 8.9208 | 1.0000 | 1.0007 | 1.0010 | 1.0016 | 1.0019 | 3.9913 |
| 2 | 8.9239 | 8.9239 | 1.0007 | 1.0010 | 1.0009 | 1.0010 | 1.0007 | 3.9909 |
| 3 | 8.9208 | 8.9216 | 1.0019 | 1.0016 | 1.0010 | 1.0007 | 1.0000 | 3.9913 |
| 4 | 8.9154 | 8.9156 | 24.3368 | 1.0019 | 1.0007 | 1.0000 | 0.9993 | 3.9893 |
| 5 | 8.9128 | 8.9130 | 8.9156 | 8.9216 | 8.9239 | 8.9208 | 8.9154 | 0.0000 |
| 6 | 8.9130 | 8.9128 | 8.9154 | 8.9208 | 8.9239 | 8.9216 | 8.9156 | 0.0000 |
| 7 | 8.9156 | 8.9154 | 0.9993 | 1.0000 | 1.0007 | 1.0019 | 24.3368 | 3.9893 |
| 8 | 0.0000 | 0.0000 | 3.9913 | 3.9913 | 3.9909 | 3.9913 | 3.9893 | 3.9917 |

**Table 4**: The calculated critical current (in units of $I_0 = 2.2623 \times 10^{-6}\ e^2\Delta/\hbar$) through a single naphthalene molecule between contacting sites i and j' calculated for $\gamma_c = 0.45\ \gamma$. Critical currents connecting two sites both labelled by a prime (or both labelled without a prime) are much smaller than $I_0$.

We performed the same calculation for larger $\gamma_c$. The molecular Josephson effect in the strong coupling limit was also studied in a recent work[46]. For stronger coupling, the critical current becomes much larger than in the weak coupling limit, which is consistent with the results of reference[46]. (The critical current increases by a factor of $10^4$ compared to the weak coupling limit.) The results for the critical currents are given in Table 5. Note that the ratio of the calculated critical currents starts to deviate from the ratios of the corresponding magic numbers.

|   | 1' | 2' | 3' | 4' | 5' | 6' | 7' | 8' |
|---|---|---|---|---|---|---|---|---|
| 1 | 8.0885 | 8.0874 | 1.0000 | 1.0031 | 1.0037 | 1.0048 | 1.0034 | 3.8543 |
| 2 | 8.0916 | 8.0916 | 1.0013 | 1.0037 | 1.0035 | 1.0037 | 1.0013 | 3.8534 |

| 3 | 8.0874 | 8.0885 | 1.0034 | 1.0048 | 1.0037 | 1.0031 | 1.0000 | 3.8543 |
| 4 | 8.0676 | 8.0679 | 18.6650 | 1.0034 | 1.0013 | 1.0000 | 0.9969 | 3.8440 |
| 5 | 8.0582 | 8.0584 | 8.0679 | 8.0885 | 8.0916 | 8.0874 | 8.0676 | 0.0000 |
| 6 | 8.0584 | 8.0582 | 8.0676 | 8.0874 | 8.0916 | 8.0885 | 8.0679 | 0.0000 |
| 7 | 8.0679 | 8.0676 | 0.9969 | 1.0000 | 1.0013 | 1.0034 | 18.6650 | 3.8440 |
| 8 | 0.0000 | 0.0000 | 3.8440 | 3.8543 | 3.8543 | 3.8543 | 3.8440 | 3.8557 |

**Table 5**: The calculated critical current (in units of $I_0 = 2.8055 \times 10^{-2} \, e^2\Delta/\hbar$) through a single naphthalene molecule between contacting sites i and i' calculated for $\gamma_c = 0.85 \, \gamma$. Critical currents connecting two sites both labelled by a prime (or both labelled without a prime) are much smaller than $I_0$.

### 4c. N-M-SS' junctions

We now examine the conductance of an N-M-SS' Andreev interferometer as a function of the superconducting phase difference $\phi_R - \phi_L$ (see Figure 5) between the left and right superconducting electrodes. When electrons go through the path connecting the normal and superconducting leads, the electronic states acquire a phase that depends on the superconducting pair potential (due to the Andreev reflection at the superconducting surface). For N-M-S junctions we have seen that the transmission amplitude related to an electron incoming from the normal electrode and reflected back as a hole is proportional to $t_{lm} \sim -M_{lm}^2 e^{-i\varphi}$. In case of N-M-SS' junctions, where there are two superconducting electrodes connected to the molecule (see Figure 5), the total transmission amplitude is a sum of the transmission amplitudes corresponding to the paths between the normal and the individual superconducting electrodes, namely $t_{l,mp} \sim -M_{lm}^2 e^{-i\varphi_L} - M_{lp}^2 e^{-i\varphi_R}$, where $l$ ($m$ and $p$) labels the site of the molecule contacted with the normal (superconducting) electrode. Hence the conductance, as a function of the superconducting phase difference can be calculated as:

$$\sigma_{l,mp}(\varphi_R - \varphi_L) \sim \left| M_{lm}^2 e^{-i\varphi_L} + M_{lp}^2 e^{-i\varphi_R} \right|^2$$
$$= M_{lm}^4 + M_{lp}^4 + 2 M_{lm}^2 M_{lp}^2 \cos(\varphi_R - \varphi_L). \tag{10}$$

The unknown coefficients from equation (10) can be dropped out by dividing it with the maximum of the conductance $\sigma_{max} = \sigma_{l,mp}(0)$. Finally, one ends up with equation (6).

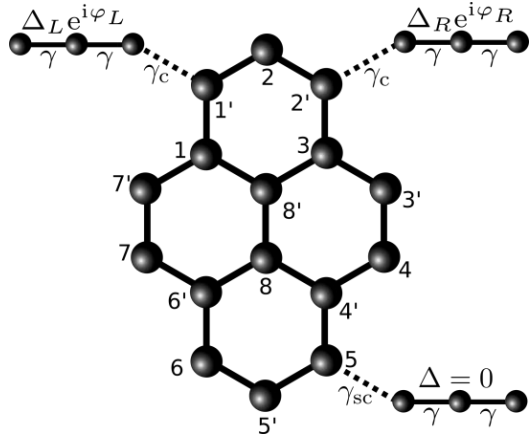

Figure 5: Andreev interferometer with pyrene molecule coupled to a normal electrode at site 5 and two superconducting contacts at site 1' and 2'

According to Equation (10), in general the conductance at the normal electrode is expected to show a periodic interference pattern as a function of the phase difference between the left and right superconducting leads. To verify our analytic expression, we compared the predictions of equation (10) to numerical tight binding simulations, see Figure 6. To calculate the conductance at the normal lead numerically, we make use of equation (10) generalized for three-terminal systems. Consequently, the Green's function, and also the normal and the Andreev reflection coefficients in equation (10) would depend on the phase difference $\phi_R - \phi_L$ of the two superconducting contacts.

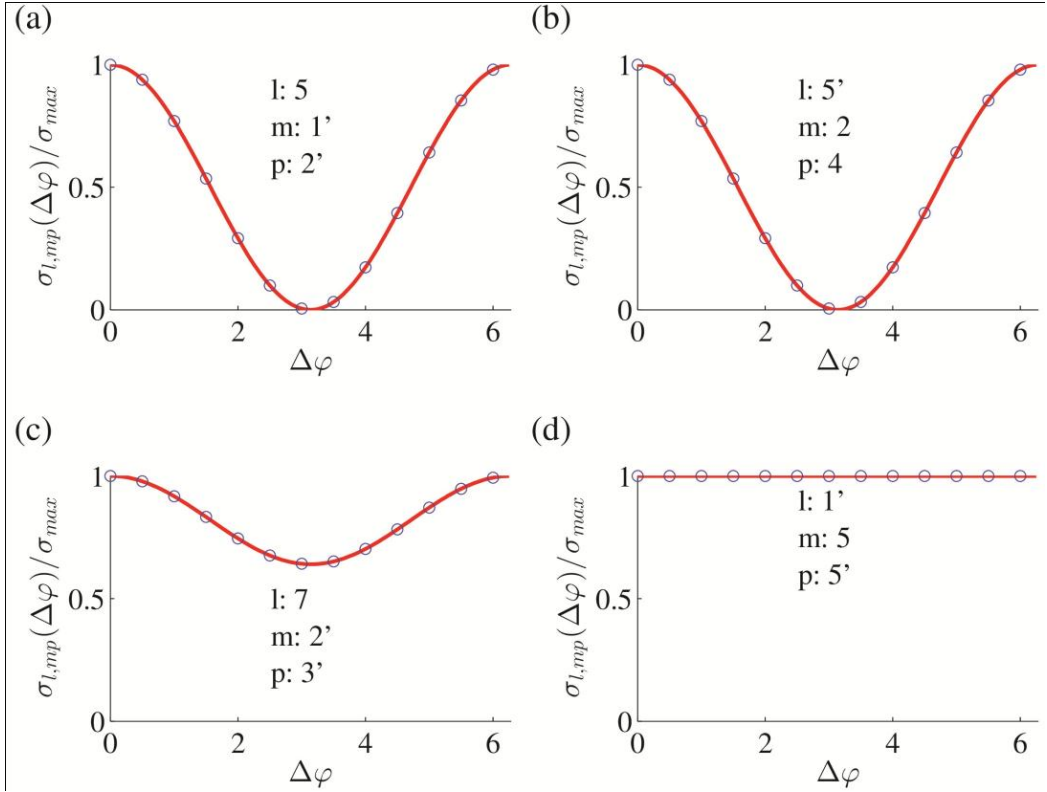

Figure 6: The calculated conductance through the normal lead at zero Fermi energy as a function of the superconducting phase difference $\Delta\phi = \phi_R - \phi_L$ at different contact positions. The solid red line represents the analytical result of equations (6) and (10), while the blue circles indicate the results of the tight binding calculations. (The positions of the left and right superconducting ($S_L$ and $S_R$) and normal (N) leads are indicated in each figure.)

One can see in Figure 6 that the interference pattern strongly depends on the position of the contacts. In agreement with the numerical results, we found that the interference pattern can only be observed if the connectivity between each pair of the normal and superconducting electrodes is finite [see Figure 6(a)-(c)], otherwise one of the interfering paths from formula (10) would be missing. Such a situation is shown in Figure 6(d), where the conductance is indeed independent of the phase difference $\phi_R - \phi_L$.

## 5. RESULTS FOR THE WIGNER DELAY TIME IN N-M-N JUNCTIONS

The Wigner delay time function was proposed by Wigner in 1955 for a single scattering channel derived from a Hermitian operator based on the scattering amplitude and then generalized by Smith in 1960 to the multichannel scattering matrices[47-49]. Delay times from different molecular orbitals can be measured experimentally, as described in [50]. More generally, within a molecular junction driven by an ac applied voltage, they are related to the off-diagonal elements of the admittance matrix, which describes the current response to such a time-varying voltage[51].

Consider a scatterer, with one-dimensional leads connected to sites $a$ and $b$, whose transmission amplitude is $t_{ab}(E) = |t_{ab}(E)| \times e^{i\theta_{ab}(E)}$. The corresponding Wigner delay time $\tau_W$ is define by $\tau_W = \hbar \tau_{ab}$, where

$$\tau_{ab} = \frac{d\theta_{ab}}{dE} \qquad (11)$$

If the scatterer is connected to single-channel current-carrying electrodes by couplings $\gamma_a$ and $\gamma_b$, then it can shown that [1]

$$t_{ab}(E) = 2i \sin k \times e^{2ik} \times \left(\frac{\gamma_a \times \gamma_b}{\gamma}\right) \times \frac{g_{ab}}{\Delta} \qquad (12)$$

where, if $H$ is the Hamiltonian describing the isolated molecular core, $g = (E - H)^{-1}$ and

$$\Delta = 1 + \frac{\gamma_a^2}{\gamma} g_{aa} e^{ik} + \frac{\gamma_b^2}{\gamma} g_{bb} e^{ik} + \frac{\gamma_a^2 \gamma_b^2}{\gamma^2} (g_{aa} g_{bb} - g_{ab} g_{ba}) e^{2ik} \qquad (13)$$

In deriving this expression, the electrodes are assumed to be one-dimensional tight-binding chains, with nearest neighbor hopping elements $-\gamma$, (where $\gamma > 0$) with a dispersion relation $E = -2\gamma \cos k$, which relates the energy $E$ of an electron travelling along the electrode to its wave vector $k$, where $0 \leq k \leq \pi$. The group velocity of such electrons within the electrodes is therefore $v = \frac{dE}{dk} = 2\gamma \sin k$.

In equation (13), $\gamma_a, \gamma_b$ are the couplings between molecule and the left and right electrodes respectively and $g_{ab}$ is the $a,b$ element of the core Green's function $g$. Since we are interested in the contribution to the delay time from the molecular core, we shall consider the 'wide band limit', where $k$ is independent of energy $E$ in the energy range of interest, between the highest occupied molecular orbital (HOMO) and lowest unoccupied molecular orbital (LUMO) of the scattering region formed by the molecule. When $H$ is real $g$ is real and therefore the delay time is obtained from the phase of the complex number $\Delta = 1 + \Delta_1 + i\Delta_2$.

In this expression, $\Delta_1 = \alpha \cos k + \beta \cos 2k$, $\Delta_2 = \alpha \sin k + \beta \sin 2k$ where, $\alpha = \frac{\gamma_a^2}{\gamma} g_{aa} + \frac{\gamma_b^2}{\gamma} g_{bb}$ and $\beta = \frac{\gamma_a^2 \gamma_b^2}{\gamma^2}(g_{aa}g_{bb} - g_{ab}g_{ba})$. Hence $\theta_{ab} = -\text{atan}(\frac{\Delta_2}{1+\Delta_1})$ and

$$\tau_{ab} = -\left[\frac{\dot{\Delta}_2(1+\Delta_1) - \dot{\Delta}_1\Delta_2}{(1+\Delta_1)^2 + \Delta_2^2}\right]$$

$$= -\left[\frac{\dot{\alpha}\sin k + \dot{\beta}\sin 2k + (\dot{\beta}\alpha - \dot{\alpha}\beta)\sin k}{(1+\alpha\cos k + \beta\cos 2k)^2 + (\alpha\sin k + \beta\sin 2k)^2}\right] \quad (14)$$

where a dot denoted a derivative with respect to $E$.

As an example, consider the mathematically simple ballistic limit, where the scatterer is a linear chain of $N$ sites coupled by nearest neighbor elements $-\gamma$. In this case, by choosing $\gamma_a = \gamma_b = \gamma$, the system reduces to a perfect linear crystal and one obtains $\theta_{ab} = k(N+1) + \frac{\pi}{2}$ and $\tau_{ab} = \frac{(N+1)dk}{dE} = \frac{N+1}{v}$, where $v = \frac{dE}{dk} = 2\gamma \sin k$ is the group velocity of a wavepacket of energy $E$. In other words, one obtains the intuitive result that the delay time is the length of the scatterer divided by the group velocity.

On the other hand, we are interested in the opposite limit of a scatterer, which is weakly coupled to the leads, such that $\frac{\gamma_a}{\gamma} \ll 1$ and $\frac{\gamma_b}{\gamma} \ll 1$ and transport is off-resonance, such that the energy $E$ lies within the HOMO-LUMO gap. (The case of on-resonance transport is discussed in appendix 1.) In this case, $\beta \ll \alpha$, and $\alpha \ll 1$ so the delay time reduces to

$$\tau_{ab} \approx -\dot{\alpha}\sin k \approx -\left(\dot{g}_{bb}\frac{\gamma_b^2}{\gamma} + \dot{g}_{aa}\frac{\gamma_a^2}{\gamma}\right)\sin k \quad (15)$$

This equation shows that the total delay time is a sum of independent times due to each contact.

$$\tau_{ab} \approx (\tau_{bb}\frac{\gamma_b{}^2}{\gamma} + \tau_{aa}\frac{\gamma_a{}^2}{\gamma})\sin k \tag{16}$$

where we have defined an intrinsic core delay time to be:

$$\tau_{aa} = -\dot{g}_{aa} \tag{17}$$

which is independent of the coupling to the leads. Since $g_{aa} = \sum_{n=1}^{N}\frac{[\psi_a(n)]^2}{E-\lambda_n}$, this yields

$$\tau_{aa} = (g^2)_{aa} = \sum_{n=1}^{N}\frac{[\psi_a(n)]^2}{(E-\lambda_n)^2} \tag{18}$$

Since the local density of states $\rho_a$ is given by

$$\rho_a = \left(-\frac{1}{\pi}\right)lim_{\eta \to 0}Im\sum_{n=1}^{N}\frac{\psi_a(n)\psi_b(n)}{E-\lambda_n+i\eta} = \eta/\pi \sum_{n=1}^{N}\frac{[\psi_a(n)]^2}{(E-\lambda_n)^2+\eta^2}.$$

This demonstrates that $\tau_{aa}$ is proportional to the local density of states at atom $a$ of the isolated molecule.

In the case where the couplings to the leads ($\gamma_a$ and $\gamma_b$) are identical, then the ratio of delay times corresponding to connectivities $a, b$ and $c, d$ is

$$\frac{\tau_{ab}}{\tau_{cd}} = \frac{\tau_{aa} + \tau_{bb}}{\tau_{cc} + \tau_{dd}} \tag{18}$$

This delay time ratio is a property of the core Green's function $g$ alone. It is interesting to note that as illustrated by all the above examples, in the weak coupling limit, the delay time is always positive.

Since $\tau_{aa} = -\dot{g}_{aa}$, where $g = (E-H)^{-1}$, $\tau_{aa}$ is obtained from the diagonal elements of $-\dot{g} = (E-H)^{-2}$, which at $E = 0$ is proportional to $M^2$, where $M$ is the magic number table of the core.

As examples, consider the graphene-like molecules shown in Figure 2, in which (a) represents a benzene ring, (b) naphthalene, (c) anthracene, (d) tetracene, (e) pentacene (f) pyrene, (g) anthanthrene and (k) azulene.

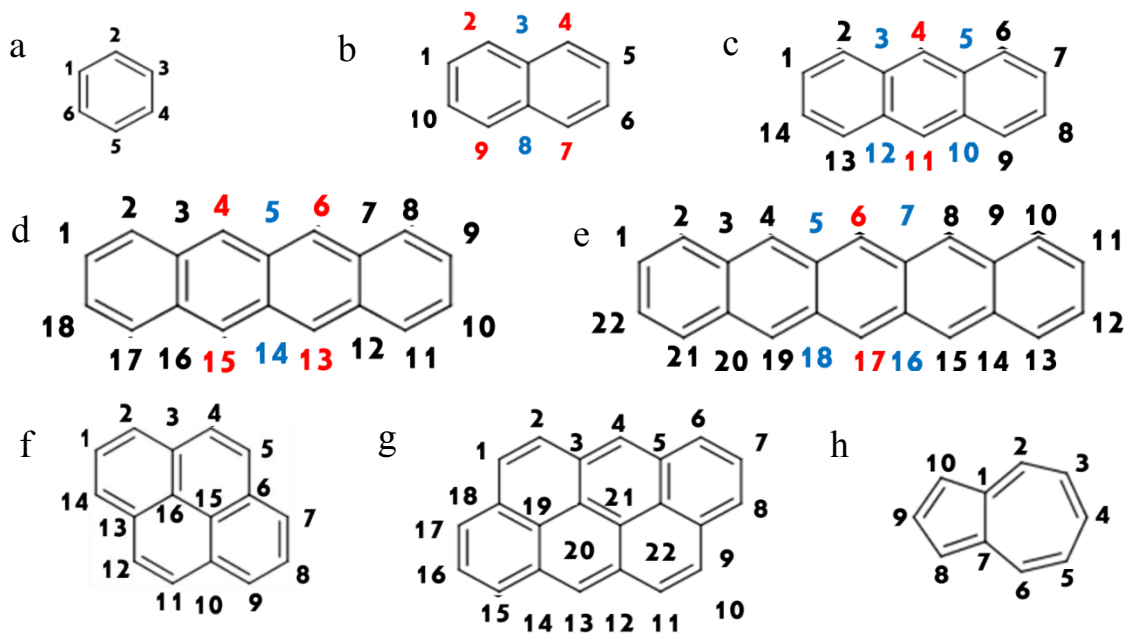

**Figure 7**: Molecular structure of substituted: a) benzene ring, b) a naphthalene, c) anthracene, d) 4-ring, e) 5-ring, f) pyrene, g) anthanthrene and h) azulene.

For the naphthalene core shown in figure 8a, figure 8b shows the Wigner delay times in the middle of HOMO-LUMO gap.

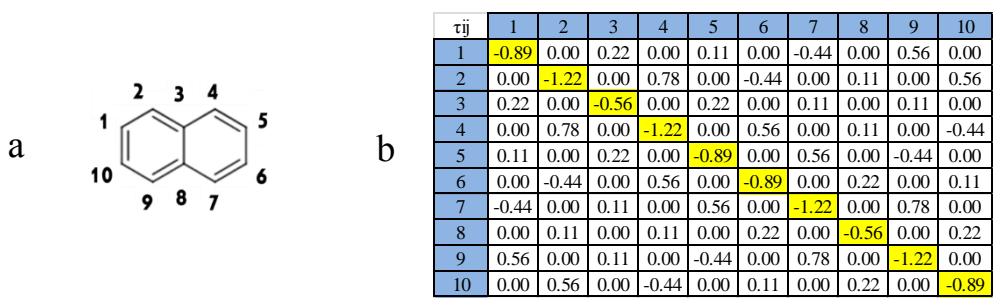

| τij | 1 | 2 | 3 | 4 | 5 | 6 | 7 | 8 | 9 | 10 |
|---|---|---|---|---|---|---|---|---|---|---|
| 1 | -0.89 | 0.00 | 0.22 | 0.00 | 0.11 | 0.00 | -0.44 | 0.00 | 0.56 | 0.00 |
| 2 | 0.00 | -1.22 | 0.00 | 0.78 | 0.00 | -0.44 | 0.00 | 0.11 | 0.00 | 0.56 |
| 3 | 0.22 | 0.00 | -0.56 | 0.00 | 0.22 | 0.00 | 0.11 | 0.00 | 0.11 | 0.00 |
| 4 | 0.00 | 0.78 | 0.00 | -1.22 | 0.00 | 0.56 | 0.00 | 0.11 | 0.00 | -0.44 |
| 5 | 0.11 | 0.00 | 0.22 | 0.00 | -0.89 | 0.00 | 0.56 | 0.00 | -0.44 | 0.00 |
| 6 | 0.00 | -0.44 | 0.00 | 0.56 | 0.00 | -0.89 | 0.00 | 0.22 | 0.00 | 0.11 |
| 7 | -0.44 | 0.00 | 0.11 | 0.00 | 0.56 | 0.00 | -1.22 | 0.00 | 0.78 | 0.00 |
| 8 | 0.00 | 0.11 | 0.00 | 0.11 | 0.00 | 0.22 | 0.00 | -0.56 | 0.00 | 0.22 |
| 9 | 0.56 | 0.00 | 0.11 | 0.00 | -0.44 | 0.00 | 0.78 | 0.00 | -1.22 | 0.00 |
| 10 | 0.00 | 0.56 | 0.00 | -0.44 | 0.00 | 0.11 | 0.00 | 0.22 | 0.00 | -0.89 |

**Figure 8:** a) Molecule structure of naphthalene. b) The $\tau_{ab}$ table of naphthalene. Note that by symmetry, there are only three distinct delay times.

To demonstrate how the Wigner delay times change with the number of the rings in the acene series a-e of Figure 7, we calculate the maximum and minimum delay times for each core as a function of the number of rings. For structures shown in Figure 7a-e, Figure 9 shows the maximum and minimum of the Wigner delay times, corresponding to the connectivities marked red and blue respectively. For example, in Figure 7b, for naphthalene, the maximum delay time is corresponds to atoms number 2, 4, 9, 7 and atoms 3 and 8 have the minimum value

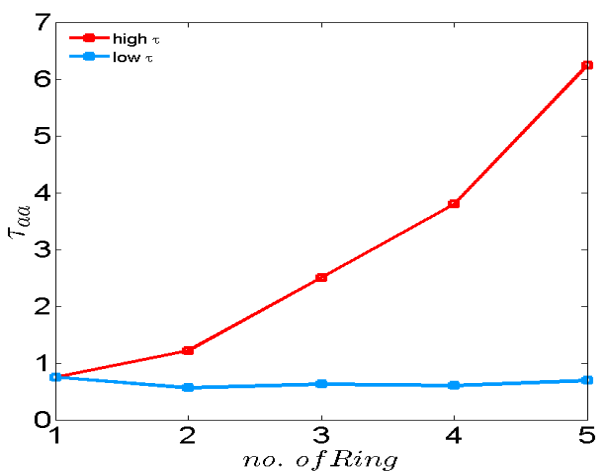

**Figure 9**: The maximum and minimum values of $\tau_{aa}$ for the acene series as a function of the number of rings.

Table 6 summarizes the minimum and maximum value of τ for different molecules.

| Molecular heart | max of τ$_{aa}$ | min of τ$_{aa}$ |
|---|---|---|
| Benzene | 0.75 | 0.75 |
| Naphthalene | 1.22 | 0.55 |
| Anthracene | 2.5 | 0.62 |
| 4_rings | 3.8 | 0.8 |
| 5_rings | 6.25 | 0.69 |
| Pyrene | 1.75 | 0.67 |
| Anthanthrene | 3.8 | 0.6 |
| Azulene | 1.97 | 1.18 |

**Table 6**: Maximum and minimum core delay times for the molecules of figure 7.

The above behavior is clearly reflected in the local density of states of the molecules, shown in Figure 10.

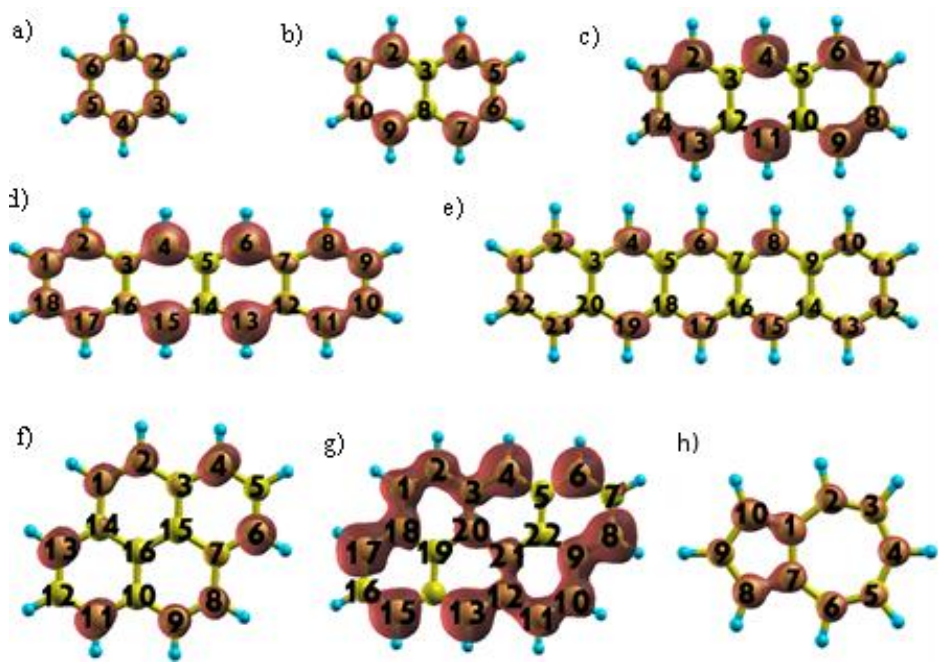

**Figure 10**: The local density of states of the molecules shown in Figure 7.

## 6. DISCUSSION

To understand how superconducting proximity effects manifest themselves in molecular-scale junctions, one needs to reconcile the vastly different energy and length scales associated with superconductivity and molecular-scale transport. For the former, typical superconducting coherence lengths are on the scale of hundreds of nanometres and energy scales on the order of meV, whereas for the latter, molecular lengths are typically a few nanometres and energies on the scale of 1-5 eV. To investigate this question experimentally, there is also a need to identify signatures of the interplay between molecular-scale transport and superconductivity, which are resilient to the sample-to-sample fluctuations arising from variability in the atomic-scale contacts between molecules and electrodes. In this article we have addressed this issue by investigating the connectivity dependence of superconducting proximity effects in various molecular structures connected to one or two superconducting contacts. We found that under certain conditions (weak coupling, locality, connectivity, mid-gap transport, phase coherence and connectivity-independent statistics) the electrical transport properties of the molecular junctions can be well described by a magic-number theory, which focusses on the connectivity between the individual sites of the molecules. For normal-molecule-superconducting junctions, for example, we find that the electrical conductance is proportional to the fourth power of their magic numbers, whereas for molecular Josephson junctions, the critical current is proportional to the square of their magic numbers. We also studied interference effects in three-terminal Andreev interferometers, where the interference pattern was driven by the superconducting phase difference. Our analytical predictions were in good agreement with the performed numerical simulations for all the studied systems.

We also investigated the connectivity dependence of Wigner delay times. At first sight, it seems unreasonable that the core Green's function and corresponding magic number table can yield information about delay times, because in the absence of a magnetic field, the core Hamiltonian and corresponding Green's function $g = (E - H)^{-1}$ are real, whereas delay times are associated with the phase of the complex transmission amplitude. Nevertheless we have demonstrated that delay time ratios can be obtained from the core Green's function or equivalently from the associated magic number tables.

**SUPPORTING INFORMATION**

Derivation of equations (4) - (6) and calculation of the Wigner delay times for on-resonance transport.

**ACKNOWLEDGEMENTS**

This work was supported by NKFIH within the Quantum Technology National Excellence Program (Project No. 2017-1.2.1-NKP-2017-00001), OTKA PD123927, K123894, K108676, K115608 and was completed

in the ELTE Institutional Excellence Program (1783-3/2018/FEKUTSRAT) supported by the Hungarian Ministry of Human Capacities. Support from the UK EPSRC is acknowledged, through grant nos. EP/N017188/1, EP/P027156/1 and EP/N03337X/1. Support from the European Commission is provided by the FET Open project 767187 – QuIET and the H2020 project Bac-To-Fuel.

# Magic number theory of superconducting proximity effects and Wigner delay times in graphene-like molecules.

*P. Rakyta[#], A. Alanazy[*], A. Kormányos[#], Z. Tajkov[+], G. Kukucska[+], J. Koltai[+], S. Sangtarash[*], H. Sadeghi[*], J. Cserti[#] and C.J. Lambert[*]*

[#]Dept. of Physics of Complex Systems, Eötvös Loránd University, Budapest, Pázmány P. s. 1/A, Hungary

[*]Dept. of Physics, Lancaster University, Lancaster, LA1 4YB, United Kingdom.

[+]Dept. of Biological Physics, Eötvös Loránd University, Budapest, Pázmány P. s. 1/A, Hungary

SUPPORTING INFORMATION

**Calculation of the Wigner delay times for on-resonance transport**

In the absence of external gating, electron transport through molecules under ambient conditions is usually off-resonance. On the other hand, if a molecule is gated such that the energy $E$ of electrons passing through the molecule is close to an energy level of the molecule, then in principle transport could be on resonance. To illustrate the properties of $\tau_{ab}$ in this limit, we now examine a number of examples, under the condition that the scatterer is weakly coupled to the leads, such that $\frac{\gamma_a}{\gamma} \ll 1$ and $\frac{\gamma_b}{\gamma} \ll 1$.

Example 1

In the limit $\beta \ll \alpha$, equation (14) of the main text reduces to

$$\tau_{ab} \approx -\left[\frac{\dot{\alpha} \sin k}{(1 + \alpha \cos k)^2 + (\alpha \sin k)^2}\right] \quad (1)$$

Example 1a. As an example, consider the case where

$$g_{ab} \approx \frac{\psi_a \psi_b}{E - \lambda_0} \quad (2)$$

In this case, $\beta = 0$ and

$$\tau_{ab} \approx \left[\frac{\Gamma_{ab}}{(E - \lambda_{ab})^2 + \Gamma_{ab}^2}\right] \quad (3)$$

In this expression, $\lambda_{ab} = \lambda_0 - \sigma$, $\sigma = \sigma_a + \sigma_b$ and $\Gamma_{ab} = \Gamma_a + \Gamma_b$, where $\sigma_a = -\frac{\gamma_a^2}{\gamma}\psi_a^2 \cos k$ and $\Gamma_a = \frac{\gamma_a^2}{\gamma}\psi_a^2 \sin k$ and similarly for $\sigma_b, \Gamma_b$. Hence on resonance, where $E = \lambda_{ab}$, the delay time reduces to $\tau_{ab} \approx 1/\Gamma_{ab}$. On the other hand, if transport is off resonance and $E$ lies close to the gap centre,

$\tau_{ab} \approx \frac{\Gamma_{ab}}{\delta^2}$, where $\delta$ is half the HOMO-LUMO gap. Since $\Gamma_{ab} \ll \delta$, this demonstrates that the 'on-resonance' delay time is much longer than the 'off-resonance' delay time.

Interestingly, the transmission coefficient $T_{ab}(E) = |t_{ab}|^2$ in this case is given by the Breit-Wigner formula:

$$T_{ab} \approx \left[\frac{4\Gamma_a \Gamma_b}{(E - \lambda_{ab})^2 + \Gamma_{ab}^2}\right] \quad (4)$$

Hence the delay time is related to the electrical conductance by

$$T_{ab}/\tau_{ab} \approx \frac{4\Gamma_a\Gamma_b}{\Gamma_a + \Gamma_b} \tag{5}$$

Example 1b

As a further example, consider the case of a Fano resonance created by a pendant orbital of energy $\epsilon$ coupled to $\psi$ by a coupling constant $\eta$, such that $\lambda_0 = \lambda_1 + \eta^2/(E - \epsilon)$. In this case, since $\lambda_0$ is energy dependent, one obtains

$$\tau_{ab} \approx \frac{\Gamma_{ab}\left(1 + \frac{\eta^2}{(E-\epsilon)^2}\right)}{\left(E - \lambda_{ab} - \frac{\eta^2}{(E-\epsilon)^2}\right)^2 + \Gamma_{ab}^2} \tag{6}$$

Near the Fano resonance, where $E \approx \epsilon$, this yields

$$\tau_{ab} \approx \frac{\Gamma_{ab}}{\eta^2} \tag{7}$$

For $\eta \ll \Gamma_{ab}$, this yields $\tau_{ab} \gg 1/\Gamma_{ab}$, which means that the electron spends a long time on the pendant orbital.

In this case,

$$T_{ab}/\tau_{ab} \approx \frac{4\Gamma_a\Gamma_b}{(\Gamma_a + \Gamma_b)\left(1 + \frac{\eta^2}{(E-\epsilon)^2}\right)} \tag{8}$$

which reflects the fact that $T_{ab}$ vanishes at the Fano resonance, where $E = \epsilon$.

**Derivation of equations (4) and (6) of the main text.**

S3

As discussed in [1], the low-bias, low-temperature electrical conductance $\sigma$ of a molecule connected to normal-metal electrode and one or more superconducting electrodes is given by

$$\sigma = \frac{4e^2}{h} R_a. \qquad (9)$$

Here $R_a$ is the Andreev reflection coefficient describing the probability that an electron from the normal electrode traverses the molecule, Andreev reflects as a hole and then exits through the normal electrode. This process is accompanied by a Cooper pair entering the superconducting electrode and therefore carries a current. The electrons and holes of such a structure are described by the Bogoliubov de Gennes Hamiltonian

$$H = \begin{pmatrix} h & \Delta e^{i\phi} \\ \Delta e^{-i\phi} & -h \end{pmatrix}. \qquad (10)$$

In this equation, $\Delta e^{i\phi}$ is the superconducting order parameter, with uniform phase $\phi$, which is non-zero in the superconducting electrode only and provides the coupling between electrons and holes. Conceptually, this doubling of the Hamiltonian, means that the pi system of the central core should be views as comprising both electronic and hole degrees of freedom, as shown in the figure below

In the absence of the electrodes, the mid-gap Greens function of the electronic system is $g^e = -h^{-1}$, whereas the Green's function of the hole system is $g^h = h^{-1}$. When the superconductor is weakly attached, the Green's function matrix element connecting the electronic pi orbital at $i$ to the hole pi orbital at $i$ is, to lowest order in the coupling, $G_{ii} = g_{ij}^h \Delta e^{-i\phi} g_{ji}^e = -(g_{ij}^e)^2 \Delta e^{-i\phi}$, where $\Delta e^{-i\phi}$ is the coupling between particles and holes mediated by the superconductor. Hence $R_a \sim |G_{ii}|^2 \sim (g_{ij}^e)^4 \sim (M_{ij})^4$

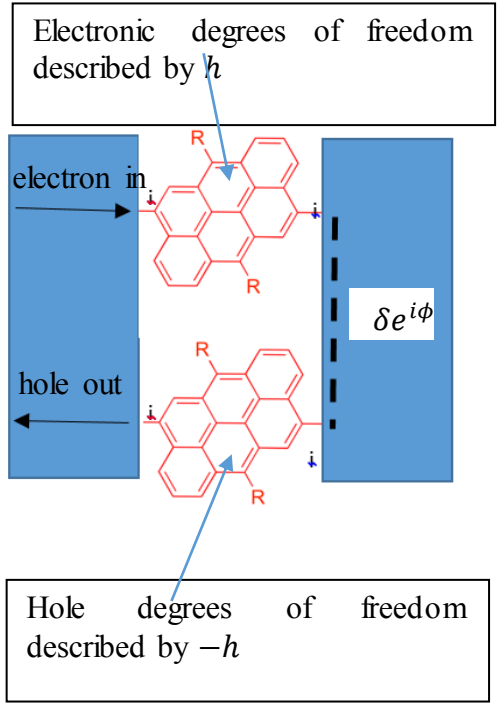

Similarly, if the molecule is weakly connected to two superconductors with order parameter phases $\phi_L$ and $\phi_R$ via pi orbitals $m$ and $p$, and to a normal electrode via pi orbital $l$,

$$G_{ll} = g^e_{lm}\Delta e^{-i\phi_L} g^h_{ml} + g^h_{lp}\Delta e^{-i\phi_R} g^h_{pl} = -\Delta[(g^e_{lm})^2 e^{-i\phi_L} + (g^h_{lp})^2 e^{-i\phi_R}]. \quad (11)$$

Hence

$$R_a \sim |G_{ii}|^2 \sim |(g^e_{lm})^2 e^{-i\phi_L} + (g^h_{lp})^2 e^{-i\phi_R}|^2 \sim |(M_{lm})^2 e^{-i\phi_L} + (M_{lp})^2 e^{-i\phi_R}|^2 \quad (12)$$

.

**Derivation of equation (5) of the main text**

According to Ref. [2], at zero temperature the critical current through a phase biased Josephson junction flowing through lead R (see Figure 4 in the main text) can be calculated by:

$$I_c^{(ij)} = \max_{\phi_L - \phi_R} \left( \frac{2}{\pi} \operatorname{Im} \int_{-\infty}^{0} dE \operatorname{Tr}\left(G_{MR}^{(ij)} \hat{I}_{MR}\right) \right). \tag{13}$$

Here $G_{MR}^{(ij)}$ and $\hat{I}_{MR}$ are the retarded Green's function matrix element and the current operator between lead R and the central molecule M. Labels $(ij)$ in this notation means, that the left superconducting lead with superconducting order parameter $\Delta_L e^{i\phi_L}$ is connected to site $i$ of the molecule and right superconducting lead with order parameter $\Delta_R e^{i\phi_R}$ to site j. The current operator is given by the coupling Hamiltonian between the molecule and the right lead, thus it is proportional to the coupling strength $\gamma_c$. (For the meaning of $\gamma_c$ see Figure 4 of the main text)

The element $G_{MR}^{(ij)}$ of the Green's function matrix $G^{(ij)}$ can be calculated via the Dyson equation

$$G^{(ij)} = \left(G_0^{-1} - V^{(ij)}\right)^{-1}, \tag{14}$$

where $G_0 = diag(G_L, G_M, G_R)$ is the Green's function of the detached system containing of the isolated central molecule $(G_M)$ and the two unattached leads ($G_L$ and $G_R$), and $V^{(ij)}$ is the coupling between the leads and the central molecule. Since the magic number theory is a weak coupling theory, we may consider $V^{(ij)}$ as a perturbation and expand the Dyson equation into series:

$$G^{(ij)} = G_0 + G_0 V^{(ij)} G_0 + G_0 V^{(ij)} G_0 V^{(ij)} G_0 \\ + G_0 V^{(ij)} G_0 V^{(ij)} G_0 V^{(ij)} G_0 + \ldots \quad . \tag{15}$$

It can be shown by straightforward calculations that the first non-vanishing contribution to the Josephson current comes from the third order term in the expansion series of the Green's function. In particular, the critical current can be given by the expression

$$I_c^{(ij)} \cong \max_{\phi_L - \phi_R} \left( \frac{2\gamma_c^4}{\pi} \operatorname{Im} \int_{-\infty}^{0} dE \left(g_{ij}^e\right)^2 \operatorname{Tr}(G_L \tau_3 G_R) \right). \tag{16}$$

In the derivation of Eq. (16) we also took advantage of the block-diagonal structure of the Green's function of the central molecule, namely $G_M = diag(g^e, g^h) = \tau_3 \otimes g^e$, where $\tau_3$ is the third Pauli matrix.

Usually the main contribution to the critical current originates from the Andreev bound states with energy located within the superconducting gap ($0 > E > -\Delta$). In this energy regime the Green's function of the central molecule can be considered to be energy independent, and the above expression can be further simplified to:

$$I_c^{(ij)} \cong \frac{2\gamma_c^4}{\pi} (g_{ij}^e)^2 \max_{\phi_L - \phi_R} \left( \mathrm{Im} \int_{-\Delta}^{0} dE\, \mathrm{Tr}(G_L \tau_3 G_R) \right). \tag{17}$$

Hence, we arrive at $I_c^{(ij)} \sim (g_{ij}^e)^2 \sim (M_{ij})^2$ that coincides with Eq. (5) in the main text.